\documentclass[12pt]{article}
\usepackage{graphicx}

\def\hybrid{\topmargin 0pt      \oddsidemargin 0pt
        \headheight 0pt \headsep 0pt
       \voffset-1cm
        \textwidth 6.25in       
       \textheight 9.5in       
        \marginparwidth 0.0in
        \parskip 5pt plus 1pt   \jot = 1.5ex}
\catcode`\@=11
\def\marginnote#1{}

\newcount\hour
\newcount\minute
\newtoks\amorpm
\hour=\time\divide\hour by60
\minute=\time{\multiply\hour by60 \global\advance\minute by-\hour}
\edef\standardtime{{\ifnum\hour<12 \global\amorpm={am}%
        \else\global\amorpm={pm}\advance\hour by-12 \fi
        \ifnum\hour=0 \hour=12 \fi
        \number\hour:\ifnum\minute<10 0\fi\number\minute\the\amorpm}}
\edef\militarytime{\number\hour:\ifnum\minute<10 0\fi\number\minute}

\def\draftlabel#1{{\@bsphack\if@filesw {\let\thepage\relax
   \xdef\@gtempa{\write\@auxout{\string
      \newlabel{#1}{{\@currentlabel}{\thepage}}}}}\@gtempa
   \if@nobreak \ifvmode\nobreak\fi\fi\fi\@esphack}
        \gdef\@eqnlabel{#1}}
\def\@eqnlabel{}
\def\@vacuum{}
\def\draftmarginnote#1{\marginpar{\raggedright\scriptsize\tt#1}}

\def\draftlabel#1{{\@bsphack\if@filesw {\let\thepage\relax
   \xdef\@gtempa{\write\@auxout{\string
      \newlabel{#1}{{\@currentlabel}{\thepage}}}}}\@gtempa
   \if@nobreak \ifvmode\nobreak\fi\fi\fi\@esphack}
        \gdef\@eqnlabel{#1}}
\def\@eqnlabel{}
\def\@vacuum{}
\def\draftmarginnote#1{\marginpar{\raggedright\scriptsize\tt#1}}

\def\draft{\oddsidemargin -.5truein
        \def\@oddfoot{\sl preliminary draft \hfil
        \rm\thepage\hfil\sl\today\quad\militarytime}
        \let\@evenfoot\@oddfoot \overfullrule 3pt
        \let\label=\draftlabel
        \let\marginnote=\draftmarginnote
   \def\@eqnnum{(\theequation)\rlap{\kern\marginparsep\tt\@eqnlabel}%
\global\let\@eqnlabel\@vacuum}  }

\def\numberbysection{\@addtoreset{equation}{section}
        \def\theequation{\thesection.\arabic{equation}}}

\def\underline#1{\relax\ifmmode\@@underline#1\else
        $\@@underline{\hbox{#1}}$\relax\fi}

\def\titlepage{\@restonecolfalse\if@twocolumn\@restonecoltrue\onecolumn
     \else \newpage \fi \thispagestyle{empty}\c@page\z@
        \def\thefootnote{\fnsymbol{footnote}} }

\def\endtitlepage{\if@restonecol\twocolumn \else  \fi
        \def\thefootnote{\arabic{footnote}}
        \setcounter{footnote}{0}}  
\relax

\hybrid

\newfont{\Bbb}{msbm10 scaled 1\@ptsize00}
\newfont{\Bbbb}{msbm7 scaled 1\@ptsize00}

\newcommand{\DDD}{\raise-1pt\hbox{$\mbox{\Bbbb D}$}}
\newcommand{\HH}{\mbox{\Bbb H}}

\newcommand{\RR}{\mbox{\Bbb R}}

\newcommand{\UUU}{\raise-1pt\hbox{$\mbox{\Bbbb U}$}}

\newcommand{\ZZ}{\mbox{\Bbb Z}}
\newcommand{\z}{\raise-1pt\hbox{$\mbox{\Bbbb Z}$}}

\def\beq{\begin{equation}}
\def\eeq{\end{equation}}
\def\p{\partial}

\begin{document}

\begin{titlepage}

\title{Elliptic parametrization of Pfaff 
integrable hierarchies in the zero dispersion limit}

\author{V.~Akhmedova\thanks{Laboratory of Mathematical Physics, 
National Research University Higher School of 
Economics, 
20 Myasnitskaya Ulitsa, Moscow 101000, Russia,
e-mail: valeria-58@yandex.ru}
\and A.~Zabrodin
\thanks{Institute of Biochemical Physics,
4 Kosygina st., Moscow 119334, Russia; ITEP, 25
B.Cheremushkinskaya, Moscow 117218, Russia and
Laboratory of Mathematical Physics,
National Research University Higher School of Economics,
20 Myasnitskaya Ulitsa,
Moscow 101000, Russia, e-mail: zabrodin@itep.ru}}

\date{December 2014}
\maketitle

\vspace{-7cm} \centerline{ \hfill ITEP-TH-43/14}\vspace{7cm}

\begin{abstract}

We show that the dispersionless limits of the Pfaff-KP 
(also known as the DKP or Pfaff lattice) and 
the Pfaff-Toda 
hierarchies admit a reformulation through elliptic functions.
In the elliptic form they look like natural elliptic 
deformations of the dispersionless KP and 2D Toda hierarchy respectively.

\end{abstract}

\end{titlepage}

\vspace{5mm}

%



\section{Introduction}

In this paper we consider dispersionless limits
of the Pfaff-KP and Pfaff-Toda hierarchies.
The aim of the paper is to present their reformulation 
in terms of elliptic functions. In this form they look like
natural ``elliptic deformations'' of the usual Kadomtsev-Petviashvili
(KP) and 2D Toda lattice (2DTL) hierarchies in the zero dispersion limit. 

The Pfaff-KP
hierarchy (also known as DKP, coupled KP, Pfaff lattice)
is one of the integrable hierarchies with 
$D_{\infty}$ symmetries introduced by 
Jimbo and Miwa in 1983 \cite{JimboMiwa}. 
Since then it emerged under different names in different contexts
\cite{HO}--\cite{Orlov}.
Its algebraic structure and some particular solutions were studied in 
\cite{Kodama,Kodama1,AKM}. The term ``Pfaff'' 
is due to the fact 
that soliton-like solutions are expressed through Pfaffians. 
In this paper we will refer to this hierarchy as the Pfaff-KP one.
Bearing certain similarities with the modified 
KP and Toda chain hierarchies,
it is essentially different and worse understood.

The 2D Pfaff-Toda hierarchy suggested in 
\cite{Willox,Takasaki09} is an extension of the 
Pfaff-KP hierarchy which relates to it in the same way as
the 2DTL relates to the KP hierarchy.
In particular, the extension $\mbox{Pfaff-KP} \longrightarrow
\mbox{Pfaff-Toda}$ implies doubling of the set of 
hierarchical times. 
Here we deal with ``real forms''
of the hierarchies which means that the KP times are assumed to be real 
while the two sets of Toda times are complex conjugate to 
each other.

The dispersionless version of the Pfaff-KP hierarchy (which 
we abbreviate as dPfaff-KP) was suggested in 
\cite{Takasaki09,Takasaki07}. In the Hirota form,
it is an infinite system
of differential equations
\beq\label{D1}
e^{D(z)D(\zeta )F}\left (1-\frac{1}{z^2\zeta^2}\,
e^{2\p_{t_0}(2\p_{t_0} + D(z) + D(\zeta ))F}\right )=
1-\frac{\p_{t_1}D(z)F -\p_{t_1}D(\zeta )F}{z-\zeta}
\eeq
\beq\label{D2}
e^{-D(z)D(\zeta )F}
\, \frac{z^2 e^{-2\p_{t_0}D(z)F}-\zeta^2 e^{-2\p_{t_0}D(\zeta )F}}{z-\zeta}
=z+\zeta -\p_{t_1}\! \Bigl (2\p_{t_0} +D(z)+D(\zeta )\Bigr )F
\eeq
for the function $F=F({\bf t})$ of the infinite number of 
(real) times
${\bf t}=\{t_0, t_1, t_2, \ldots \}$,
where 
\beq\label{D3}
D(z)=\sum_{k\geq 1}\frac{z^{-k}}{k}\, \p_{t_k}.
\eeq
The function $F$ is a dispersionless analogue of the tau-function.
The differential equations are obtained by expanding 
equations (\ref{D1}), (\ref{D2}) in powers of $z$, $\zeta$.
For example, the first two equations of the hierarchy are
\beq\label{simp1}
\left \{ 
\begin{array}{l}
6F_{11}^2 +3F_{22}-4F_{13}=12e^{4F_{00}}
\\ \\
2F_{03}+4F_{01}^3 +6 F_{01}F_{11}-6F_{01}F_{02}=3F_{12}.
\end{array}
\right.
\eeq
We use the short-hand notation
$F_{mn}\equiv \p_{t_m}\p_{t_n}F$.

The dispersionless version of the Pfaff-Toda hierarchy (dPfaff-Toda)
\cite{Takasaki09} is written for a function $F$ of the doubly-infinite
set of
times $\{\ldots , \bar t_2, 
\bar t_1, r,s,  t_1, t_2, \ldots \}$. 
Since the different hierarchies are never mixed in this paper,
we keep the same notation $F$ for the dispersionless tau-function.
The real form of the hierarchy, which we will be dealt with,
implies that $\bar t_k$ is complex conjugate to $t_k$, $s$ is real
and $r$ is purely imaginary. 
The basic equations are as follows:
\beq\label{pft1}
e^{D(z)D(\zeta)F}\left (1-\frac{1}{z\zeta}\,
e^{\p_s
\bigl (\p_s +\p_r+D(z)+D(\zeta)\bigr )F}\right )=
\frac{ze^{-\p_r D(z)F}-\zeta
e^{-\p_r D(\zeta )F}}{z-\zeta},
\eeq
\beq\label{pft1a}
e^{\bar D(\bar z)\bar D(\bar \zeta)F}
\left (1-\frac{1}{\bar z\bar \zeta}\,
e^{\p_s
\bigl (\p_s -\p_r +\bar D(\bar z)+\bar D(\bar 
\zeta)\bigr )F}\right )=
\frac{\bar ze^{\p_r \bar D(\bar z)F}-\bar \zeta
e^{\p_r \bar D(\bar \zeta )F}}{\bar z-\bar \zeta},
\eeq
\beq\label{pft2}
e^{D(z)D(\zeta)F}\left (1-\frac{1}{z\zeta}\,
e^{\p_r \bigl (\p_s +\p_r +D(z)+D(\zeta)\bigr )F}\right )=
\frac{ze^{-\p_s D(z)F}-\zeta
e^{-\p_s D(\zeta )F}}{z-\zeta},
\eeq
\beq\label{pft2a}
e^{\bar D(\bar z)\bar D(\bar \zeta)F}
\left (1-\frac{1}{\bar z\bar \zeta}\,
e^{-\p_r \bigl (\p_s-\p_r+
\bar D(\bar z)+\bar D(\bar \zeta)\bigr )F}\right )=
\frac{\bar ze^{-\p_s \bar D(\bar z)F}-\bar \zeta
e^{-\p_s \bar D(\bar \zeta )F}}{\bar z-\bar \zeta},
\eeq
\beq\label{pft3}
e^{-D(z)\bar D(\bar \zeta)F}\left (1-\frac{1}{z\bar \zeta}\,
e^{\p_r \bigl (\p_r +D(z)-
\bar D(\bar \zeta)\bigr )F}\right )=1-\frac{1}{z\bar \zeta}\,
e^{\p_s \bigl (\p_s +D(z)+
\bar D(\bar \zeta)\bigr )F},
\eeq
\beq\label{pft4}
e^{-\bigl (\p_s+\p_r +D(z)\bigr )
\bar D(\bar \zeta)F}-1=
\frac{z}{\bar \zeta}\,
e^{-\p_r \bigl (\p_s +D(z)+\bar D(\bar \zeta)\bigr )F}
\! \left (e^{-\bigl (\p_s-\p_r +\bar D(\bar \zeta)\bigr ) D(z)F}-1\right ).
\eeq
Here $\displaystyle{\bar D(\bar z)=\sum_{k\geq 1}\frac{\bar z^{-k}}{k}\, 
\p_{\bar t_k}}$ is the complex conjugate counterpart of the 
differential operator (\ref{D3}). 
Note that equations (\ref{pft1a}), (\ref{pft2a}) are obtained from,
respectively, (\ref{pft1}), (\ref{pft2}) by applying the 
``bar-operation'' $D\to \bar D$, $z\to \bar z$, $\zeta \to \bar \zeta$,
$t_k\to \bar t_k$, $s\to \bar s=s$, $r\to \bar r=-r$ which can be treated as complex conjugation 
provided the function $F$ is real. We see that each equation has 
a ``bar-counterpart''. 
At the same time, the other two 
equations,
(\ref{pft3}) and (\ref{pft4}), are real,
i.e. they do not change under the complex conjugation.
From now on we will not always write explicitly
the conjugates of complex equations keeping in mind that they 
hold simultaneously.
In what follows it will be more convenient to introduce 
the complex conjugate ``0th times''
$t_0=s+r$, $\bar t_0=s-r$, so $\p_{t_0}=\frac{1}{2}(\p_s +\p_r)$,
$\p_{\bar t_0}=\frac{1}{2}(\p_s -\p_r)$.

The differential equations are obtained by expanding 
(\ref{pft1})--(\ref{pft4}) in powers of $z$, $\zeta$, 
$\bar z$, $\bar \zeta$. The two simplest equations of the hierarchy
are
\beq\label{simp2}
\left \{ 
\begin{array}{l}
e^{F_{00}}F_{0\bar 1}=e^{F_{\bar 0\bar 0}}F_{\bar 0 1},
\\ \\
F_{1\bar 1}=2\, e^{F_{00}+F_{\bar 0\bar 0}}\sinh \bigl (2F_{0\bar 0}\bigr ).
\end{array}\right.
\eeq
Here $F_{mn}\equiv \p_{t_m}\p_{t_n}F$, 
$F_{m\bar n}\equiv \p_{t_m}\p_{\bar t_{n}}F$, 
$F_{\bar m\bar n}\equiv \p_{\bar t_{m}}\p_{\bar  t_{n}}F$.

In this work we show that the Pfaff-type hierarchies admit 
a nice reformulation in terms of elliptic functions 
(or Jacobi theta functions). 
After this reformulation, the number of independent 
equations gets reduced and
somewhat unsightly looking equations (\ref{D1}), (\ref{D2})
and especially (\ref{pft1})--(\ref{pft4})
assume compact
and suggestive forms which look like natural elliptic deformations
of the dispersionless KP 
(or modified KP) and 2DTL  hierarchies (see respectively
(\ref{dkp}), (\ref{dmkp}) and (\ref{dtoda}) below in Section 4).
Note that in the elliptic parametrization, 
the modular parameter $\tau$ is a dynamical variable.
This feature suggests some 
similarities with the genus 1 Whitham equations \cite{Krichever}
and the integrable structures behind boundary value 
problems in 
doubly-connected domains in the plane \cite{KMZ05}.

The elliptic form of the dPfaff-KP hierarchy was 
obtained in our previous work \cite{AZ14}. 
This result is reviewed in Section 2. 
In Section 3 it is extended to  
the dPfaff-Toda case. In Section 4 we compare 
the Pfaff-type hierarchies with the more familiar ones.

\section{The dispersionless Pfaff-KP hierarchy}

\paragraph{Algebraic formulation.}
Here we deal with the set of real times 
${\bf t}=\{t_0, t_1, t_2, \ldots \}$. 
In what follows we use the differential operator
\beq\label{E5}
\nabla (z)=\p_{t_0}+ D(z)
\eeq
which appears to be more convenient than $D(z)$. 
Introducing the auxiliary functions
\beq\label{D4}
p(z)=z-\p_{t_1} \nabla (z)F,
\qquad
w(z)=z^2 e^{-2\p_{t_0}\nabla (z)F},
\eeq
we can rewrite equations (\ref{D1}), (\ref{D2}) in a more compact form
\beq\label{D1a}
e^{D(z)D(\zeta )F}\left (1-\frac{1}{w(z)w(\zeta )}\right )=
\frac{p(z)-p(\zeta )}{z-\zeta}\,,
\eeq
\beq\label{D2a}
e^{-D(z)D(\zeta )F + 2\p_{t_0}^2F}
\, \, \frac{w(z)-w(\zeta )}{z-\zeta}=p(z)+p(\zeta ).
\eeq
Multiplying them, we get the relation
$$
p^2(z)-e^{2F_{00}}\Bigl (w(z)+w^{-1}(z)\Bigr )=
p^2(\zeta )-e^{2F_{00}}\Bigl (w(\zeta )+w^{-1}(\zeta )\Bigr )
$$
which states that the combination 
$p^2(z)-e^{2F_{00}}\Bigl (w(z)+w^{-1}(z)\Bigr )$
does not depend on $z$. The limit $z\to \infty$ allows one 
to express this quantity
through derivatives of the function $F$. 
As a result we find that $p(z), w(z)$ satisfy the
algebraic equation \cite{Takasaki09}
\beq\label{D5}
p^2(z)=\mathsf{r}^2 \Bigl (w(z)+w^{-1}(z)\Bigr )- \mathsf{v} ,
\eeq
where
$\mathsf{r}=e^{F_{00}}$,
$\mathsf{v}=2F_{11} +F^2_{01}-F_{02}$ are real parameters.
This equation defines an elliptic curve, with
$p$, $w$ being algebraic functions on it.  The local 
parameter around $\infty$ is $z^{-1}$. 
As is seen from (\ref{D4}), 
the functions $p$ and $w$ have respectively a simple 
and a double pole at infinity.

\paragraph{Elliptic formulation.}
A natural further step is to  
uniformize the elliptic curve (\ref{D5})
using the elliptic functions $\mbox{sn}$, 
$\mbox{cn}$, $\mbox{dn}$ or the 
theta-functions\footnote{Their definition and basic properties 
are listed in the appendix.
Below we will often write simply $\theta_a(u)$ if this
does not cause confusion.} 
$\theta_a (u)=\theta_a (u|\tau )$ 
($a=1,2,3,4$).
It can be done in different ways. 
Note first that given $\mathsf{r}$, $\mathsf{v}$, 
the modular parameter $\tau \in \HH$ ($\HH$ is the 
upper half-plane) is not uniquely defined
because of possible modular transformations. 
The reality 
of the coefficients  $\mathsf{r}^2$, $\mathsf{v}$ 
implies certain restrictions on possible 
values of $\tau$.
In the standard fundamental domain 
$\Bigl \{ \tau \in \HH \Bigm | 
|\mathrm{Re}\, \tau |\leq \frac{1}{2}, |\tau |\geq 1\Bigr \}$
possible values of $\tau$ are as follows: 
a) $\tau =it$, $t\geq 1$, b)  $\tau =\frac{1}{2}+it$, $t\geq 
\sqrt{3}/2$,
c) $\tau =e^{i\rho}$, $\frac{\pi}{3}\leq \rho \leq \frac{2\pi}{3}$.

In what follows we will consider purely imaginary $\tau$ (case a))
and choose the uniformization suggested in \cite{AZ14}:
\beq\label{E1}
w(z)=\frac{\theta_4^2(u(z))}{\theta_1^2(u(z))}\,, \qquad
p(z)=\gamma \, \theta_4^2(0)\, \frac{\theta_2(u(z))\,
\theta_3(u(z))}{\theta_1(u(z))\, \theta_4(u(z))}\,, 
\eeq
with $\mathsf{r}$, $\mathsf{v}$ given by
\beq\label{E2}
\mathsf{r}=\gamma\, \theta_2(0)\, \theta_3(0)\,, \qquad
\mathsf{v}=\gamma^2 \Bigl ( \theta_2^4(0)+\theta_3^4(0)\Bigr ).
\eeq
One can check that 
the equation of the curve becomes equivalent to
the identity
\beq\label{E222}
\theta_4^4(0)\, \frac{\theta_2^2(u)\,
\theta_3^2(u)}{\theta_1^2(u)\, \theta_4^2(u)}=
\theta_2^2(0)\theta_3^2(0)\! \left (\frac{\theta_4^2(u)}{\theta_1^2(u)}+
\frac{\theta_1^2(u)}{\theta_4^2(u)}\right )-
\Bigl (\theta_2^4(0)+\theta_3^4(0)\Bigr ),
\eeq
which can be proved in the standard way 
by comparing analytical properties 
of the both sides.
The $z$-independent factor $\gamma \in \RR$ in
(\ref{E1}), (\ref{E2}) is, at this stage,
an arbitrary 
parameter. As is shown below, it is a 
dynamical variable, as well as the modular parameter $\tau$:
$\gamma =\gamma ({\bf t})$, $\tau =\tau ({\bf t})$. 
The function $u$ in (\ref{E1}) depends on $z$ and on all times:
$u(z)=u(z, {\bf t})$. 
Equations (\ref{D4}) show that the functions 
$w(z), p(z)$ take real values for real $z$. 
Taking this into account, it is convenient to 
normalize $u(z)$ by the condition $u(\infty )=0$, with 
the expansion around $\infty$ being of the form
\beq\label{E3}
u(z, {\bf t})=\frac{c_1({\bf t})}{z}+\frac{c_2({\bf t})}{z^2}+\ldots \, ,
\quad c_i\in \RR .
\eeq

After the uniformization equations (\ref{D1a}) and (\ref{D2a})
become identical. 
Let us take, for example, equation (\ref{D2a}) and write it as
$$
(z_1^{-1}-z_2^{-1})e^{\nabla_1 \nabla_2 F}=-\,
\frac{w_1-w_2}{p_1-p_2}\, \frac{\mathsf{r}}{\sqrt{w_1w_2}}\, ,
$$
where $\nabla_i=\nabla (z_i)$, $p_i=p(z_i)$, etc. 
The identity
$$
\frac{w_1-w_2}{p_1+p_2}=-\frac{1}{\gamma \,
\theta_2(0)\theta_3(0)}\,\,
\frac{\theta_4(u_1)\theta_4(u_2)}{\theta_1(u_1)\theta_1(u_2)}\,\,
\frac{\theta_1(u_{12})}{\theta_4(u_{12})}
$$
(here and below $u_i\equiv u(z_i)$, $u_{ik}\equiv 
u_i-u_k$) allows us to transform the right hand side to 
a very simple form:
\beq\label{E4}
\mbox{\fbox{$\displaystyle{\phantom{\int\limits^{A}_{B}}
\left (z_1^{-1}-z_2^{-1}\right ) e^{\nabla (z_1) \nabla (z_2)F}
=\frac{\theta_1 \bigl (u(z_1)\! -\! u(z_2)\bigr )}{\theta_4
\bigl (u(z_1)\! -\! u(z_2)\bigr )}\,.
\phantom{\int ^{A}_{B}}}$}}
\eeq
This equation encodes the dispersionless Pfaff-KP hierarchy.
Note that the limit $z_2\to \infty$ in (\ref{E4}) 
gives the definition of the 
function $u(z)$ equivalent to the first formula in (\ref{E1}):
\beq\label{E6}
e^{\p_{t_0}\nabla (z)F}=z\,
\frac{\theta_1(u(z))}{\theta_4(u(z))}.
\eeq
The $z\to \infty$ limit of this equation yields
$e^{F_{00}}=\mathsf{r}=\pi c_1 \theta_2(0)\theta_3(0)$,
hence
\beq\label{E9}
c_1({\bf t})=\frac{\gamma ({\bf t})}{\pi}, 
\qquad
\gamma ({\bf t})= \frac{e^{F_{00}}}{\theta_2(0|\tau )\theta_3(0|\tau )}.
\eeq
In addition, we see from (\ref{E2}) that
\beq\label{E7}
\frac{\mathsf{v}}{\mathsf{r}^2}=\, e^{-2F_{00}}\! \left (
2F_{11} \! +\! F^2_{01} \! -\! F_{02}\right )\, 
=\,
\frac{\theta_2^2(0|\tau )}{\theta_3^2(0|\tau )}+
\frac{\theta_3^2(0|\tau )}{\theta_2^2(0|\tau )}\,.
\eeq
This relation makes it clear that the modular parameter
$\tau$ is expressed through second order partial derivatives 
of the function $F$. Due to (\ref{E9}) the same is true for 
$c_1$ and $\gamma$.

To give yet another instructive form of equation (\ref{E4}),
it is convenient to introduce the function
\beq\label{E10}
S(u| \, \tau ):=\log \frac{\theta_1(u |\tau )}{\theta_4(u |\tau )}.
\eeq
It has the (quasi)periodicity properties
$S(u+1|\tau )=S(u|\tau )+i\pi$, 
$S(u+\tau |\tau )=S(u|\tau )$. The derivative of this function
$S'(u)=\p_u S(u|\tau)$ is given by
\beq\label{Sprime}
S'(u)=\pi \theta_4^2(0)\, 
\frac{\theta_2(u)\theta_3(u)}{\theta_1(u)\theta_4(u)}.
\eeq
This formula can be easily proved, with the help of identity
(\ref{theta1prime}) from the appendix, 
by comparing analytical properties
of the both sides.
For the needs of the next section we note here that
the identity (\ref{E222}) can be read as a 
non-linear differential equation for
the function $S$:
\beq\label{diffS}
\left (\frac{S'(u)}{\pi \theta_2(0)\theta_3(0)}\right )^2=
2\cosh \bigl (2S(u)\bigl )-
\frac{\theta_2^2(0)}{\theta_3^2(0)}-\frac{\theta_3^2(0)}{\theta_2^2(0)}.
\eeq
Let us take 
logarithms and apply $\p_{t_0}$ to both sides of (\ref{E4}). 
In terms of the function $S(u)$, the equation reads
\beq\label{E11}
\nabla (z_1) S\Bigl (u(z_2)|  \tau\Bigr ) =
\p_{t_0}S\Bigl (u(z_1) \! -\! u(z_2)| \tau \Bigr ).
\eeq
In particular, this equation means that the left hand side 
is symmetric with respect to the permutation $z_1 \leftrightarrow z_2$:
$\nabla (z_1) S\Bigl (u(z_2)| \tau\Bigr ) =
\nabla (z_2) S\Bigl (u(z_1)| \tau\Bigr )$. This symmetry 
is a manifestation of integrability.
In the limit $z_2\to \infty$ equation (\ref{E11}) gives:
\beq\label{E12}
\nabla (z)\log \mathsf{r}
= \p_{t_0} S\Bigl (u(z)\bigr | \tau\Bigr ).
\eeq
To connect this with the algebraic formulation, we note that
\beq\label{E13}
S(u(z)|\tau )= -\frac{1}{2}\, \log w(z), \qquad
c_1 S'(u(z)|\tau )= p(z).
\eeq
These formulas directly
follow from the definitions and from
(\ref{Sprime}).

\section{The dispersionless Pfaff-Toda hierarchy}

\paragraph{Algebraic formulation.}
Now the set of times is ${\bf t}=
\{\ldots , \bar t_2, \bar t_1, \bar t_0,
t_0, t_1, t_2, \ldots \}$. Accordingly, the operator (\ref{E5}) 
acquires the ``bar-counterpart'' $\bar \nabla (\bar z)=
\p_{\bar t_0}+\bar D(\bar z)$. From now on we will work with 
the times $t_0$, $\bar t_0$ instead of $s,r$.

Introducing the auxiliary functions
\beq\label{T1}
\begin{array}{l}
P(z)=ze^{-(\p_{t_0}+\p_{\bar t_0})\nabla (z)F}, \qquad
W(z)=ze^{-(\p_{t_0}-\p_{\bar t_0})\nabla (z)F},
\\ \\
\bar P(z)=ze^{-(\p_{t_0}+\p_{\bar t_0})\bar \nabla (z)F}, \qquad
\bar W(z)=ze^{(\p_{t_0}-\p_{\bar t_0})\bar \nabla (z)F},
\end{array}
\eeq
we can rewrite equations (\ref{pft1})--(\ref{pft4}) in a more
compact form
\beq\label{T2}
\begin{array}{l}
\displaystyle{
e^{D(z)D(\zeta )F}\left (1-\frac{1}{P(z)P(\zeta )}\right )=
\frac{W(z)-W(\zeta )}{z-\zeta}\, e^{(\p_{t_0}-\p_{\bar t_0})\p_{t_0}F}}
\\ \\
\displaystyle{
e^{D(z)D(\zeta )F}\left (1-\frac{1}{W(z)W(\zeta )}\right )=
\frac{P(z)-P(\zeta )}{z-\zeta}\, e^{(\p_{t_0}+\p_{\bar t_0})\p_{t_0}F}}
\\ \\
\displaystyle{
e^{D(z)\bar D(\bar \zeta )F}\left (1-\frac{1}{P(z)
\overline{P(\zeta)}}\right )=
1-\frac{1}{W(z)\overline{W(\zeta)}}}
\\ \\
\displaystyle{
e^{D(z)\bar D(\bar \zeta)F}\left (W(z)-\overline{W(\zeta)}\right )=
\left (P(z)-\overline{P(\zeta)}\right )
e^{2\p_{t_0}\p_{\bar t_0} F }},
\end{array}
\eeq
where $\overline{P(\zeta)}:=\bar P(\bar z)$,
$\overline{W(\zeta)}:=\bar W(\bar z)$.
Dividing the first equation by the second one, we get the relation
$$
W(z)+W^{-1}(z)-e^{2\p_{t_0}\p_{\bar t_0}F}
\! \left (P(z)+P^{-1}(z)\right )=
W(\zeta )+W^{-1}(\zeta)-
e^{2\p_{t_0}\p_{\bar t_0}F}
\! \left (P(\zeta)+P^{-1}(\zeta)\right )
$$
which states that the combination 
$W(z)+W^{-1}(z)-e^{2\p_{t_0}\p_{\bar t_0}F}\! 
\left (P(z)+P^{-1}(z)\right ):=C$ does not depend on $z$. 
The limit $z\to \infty$ allows one to express the constant $C$
through derivatives of the function $F$:
$
C=2e^{-(\p_{t_0}-\p_{\bar t_0})\p_{t_0}F}\p_{\bar t_0}\p_{t_1}F
$.
Dividing the third equation in (\ref{T2}) by the fourth one,
we get a relation which states that $C$ is real, i.e.,
$e^{\p_{t_0}^2 F}\p_{t_0}\p_{\bar t_1}F=e^{\p_{\bar t_0}^2F}
\p_{\bar t_0}\p_{t_1}F$. This is the first equation in
(\ref{simp2}).
As a result, we find that $P(z), W(z)$ satisfy the algebraic
equation \cite{Takasaki09}
\beq\label{T3}
W(z)+W^{-1}(z)-R^2\left ( P(z)+P^{-1}(z)\right )=C,
\eeq
with the real coefficients
\beq\label{T4}
R^2=e^{2F_{0\bar 0}}, \qquad
C= 2e^{F_{0\bar 0}-F_{00}}F_{\bar 01}.
\eeq
The functions $\bar P$, $\bar W$ satisfy the same equation.
Like in the case of the dPfaff-KP hierarchy, this equation
defines an elliptic curve, with $P$ and $W$ being algebraic 
functions on it and $z^{-1}$ the local 
parameter around $\infty$. 
As is seen from (\ref{T1}), 
both $P$ and $W$ have a simple pole at infinity.

In what follows it is more convenient to work with 
the functions 
\beq\label{T5}
f(z)=\sqrt{P(z)W(z)}=ze^{-\p_{t_0}\nabla (z)F}, \qquad
g(z)=\sqrt{P(z)/W(z)}=e^{-\p_{\bar t_0}\nabla (z)F}.
\eeq
The function $f$ has a simple pole at $\infty$ while $g$ is regular
there.
Their complex conjugates are 
$\overline{f(z)}=\bar f(\bar z)=
\bar ze^{-\p_{\bar t_0}\bar \nabla (\bar z)F}$,
$\overline{g(z)}=\bar g(\bar z)=
e^{-\p_{t_0}\bar \nabla (\bar z)F}$.
In these terms the equation of the elliptic curve reads
\beq\label{T6}
R^2(f^2g^2 +1)+Cfg=f^2+g^2.
\eeq
Note the symmetry $f\leftrightarrow g$. The functions 
$\bar f(z)$, $\bar g(z)$ obey the same equation. 

\paragraph{Elliptic formulation.} The uniformization of the curve
(\ref{T6}) in terms of the 
theta functions $\theta_a(u)=\theta_a(u|\tau )$ can be chosen in the form
\beq\label{ET1}
f(z)=\frac{\theta_4(u(z))}{\theta_1(u(z))}, \qquad
g(z)=\frac{\theta_4(u(z)+\eta )}{\theta_1(u(z)+\eta )}.
\eeq
Here $u(z)=u(z, {\bf t})$ has the same expansion (\ref{E3})
around $\infty$ but the coefficients are complex and there is also 
the series $\overline{u(z)}=
\bar u(\bar z)=\bar u(\bar z, {\bf t})$ with 
conjugate coefficients:
\beq\label{E3a}
u(z, {\bf t})=
\frac{c_1({\bf t})}{z}+
\frac{c_2({\bf t})}{z^2}+\ldots \, ,
\qquad
\bar u(z, {\bf t})=\frac{\overline{c}_1({\bf t})}{z}+
\frac{\overline{c}_2({\bf t})}{z^2}+\ldots \, .
\eeq
The parameter
$\eta$ is 
a dynamical variable as well as the modular parameter $\tau$:
$\eta = \eta ({\bf t})$, $\tau = \tau ({\bf t})$. 
Plugging (\ref{ET1}) 
into the equation of the curve, one can see that it converts into 
identity if
\beq\label{ET2}
R=\frac{\theta_1(\eta)}{\theta_4(\eta)}, \qquad
C=2\, \frac{\theta_4^2(0)\, 
\theta_2(\eta)\, \theta_3(\eta)}{\theta_4^2(\eta)\, 
\theta_2(0)\, \theta_3(0)}.
\eeq
We assume that $\eta$ is real 
and $\tau$ is purely imaginary. This is consistent with 
reality of $R$ and $C$.

After the uniformization only two equations in (\ref{T2})
remain independent (say, the first and the third one). 
Our next task is to represent them in the elliptic form.
Let us first rewrite them as
$$
(z_1^{-1}-z_2^{-1})e^{\nabla_1\nabla_2 F}
=R^{-1}g_1g_2 \, \frac{W_1-W_2}{1-P_1P_2},
$$
$$
e^{\nabla_1\bar \nabla_2 F}
=R^{-1}g_1\bar g_2 \, \frac{1-W_1\bar W_2}{1-P_1\bar P_2},
$$
where $\nabla_i =\nabla (z_i)$, $\bar \nabla_i =\bar \nabla (\bar z_i)$,
$g_i=g(z_i)$, etc. The identities
$$
\frac{W_1-W_2}{1-P_1P_2}\, =\,\, \frac{\theta_1(\eta)}{\theta_4(\eta)}\,
\frac{\theta_1(u_1+\eta)\, \theta_1(u_2+\eta)}{\theta_4(u_1+\eta)\,
\theta_4(u_2+\eta)}\, \cdot \frac{\theta_1(u_1-u_2)}{\theta_4(u_1-u_2)},
$$
$$
\frac{1-W_1\bar W_2}{1-P_1\bar P_2}\, =
\,\, \frac{\theta_1(\eta)}{\theta_4(\eta)}\,
\frac{\theta_1(u_1+\eta)\, \theta_1(\bar u_2+\eta)}{\theta_4(u_1+\eta)\,
\theta_4(\bar u_2+\eta)}\, \cdot
\frac{\theta_1(u_1+\bar u_2+\eta)}{\theta_4(u_1+\bar u_2+\eta)}
$$
allow one to represent the equations in the form
\beq\label{ET3}
\mbox{\fbox{$\displaystyle{\phantom{\int\limits^{A}_{B}}
\begin{array}{rll}
(z_1^{-1}-z_2^{-1})\, e^{\nabla (z_1)\nabla (z_2)F}&=&\displaystyle{
\frac{\theta_1 (u(z_1)-u(z_2))}{\theta_4 (u(z_1)-u(z_2))}}
\\ &&\\
e^{\nabla (z_1)\bar \nabla (\bar z_2)F}&=&\displaystyle{
\frac{\theta_1 (u(z_1)+\bar u(z_2)+\eta )}{\theta_4 
(u(z_1)+\bar u( z_2)+\eta )}}
\\ &&\\
(z_1^{-1}-z_2^{-1})\, e^{\bar \nabla (z_1)\bar 
\nabla (z_2)F}&=&\displaystyle{
\frac{\theta_1 (\bar u(z_1)-
\bar u(z_2))}{\theta_4 (\bar u(z_1)-\bar u(z_2))}}
\end{array}
\phantom{\int ^{A}_{B}}}$}}
\eeq
The first equation is the same as (\ref{E4}). This means that a ``half''
of the dispersionless Pfaff-Toda hierarchy (with fixed bar-times)
coincides with the Pfaff-KP one. This fact can not be so 
transparently seen in the algebraic formulation. 
The third equation is the bar-version of the first one.
It represents another copy of the 
dPfaff-KP hierarchy, now with respect to the bar-times 
$\bar t_k$ with fixed $t_k$'s. The second equation 
contains mixed derivatives with respect to the times 
$\{t_k\}$ and $\{\bar t_k\}$ and thus it 
couples the two hierarchies into the more general one.
This equation is 
invariant under complex conjugation.

The $z_2\to \infty$ limits of equations (\ref{ET3}) yield:
\beq\label{ET4}
e^{\p_{t_0}\nabla (z)F}=z\, \frac{\theta_1(u(z))}{\theta_4(u(z))},
\qquad
e^{\p_{\bar t_0}\nabla (z)F}\, =\, 
\frac{\theta_1(u(z)+\eta )}{\theta_4(u(z)+\eta )},
\eeq
which are nothing else than the expressions for the functions
$f$ and $g$ (\ref{ET1}) combined with their definition
(\ref{T5}). The further $z\to \infty$ expansion of these relations
gives
$
e^{F_{00}}=\pi c_1\theta_2(0)\theta_3(0)
$
from the leading terms of the first one and
$F_{\bar 01}=c_1 S'(\eta)$ from the 
$O(z^{-1})$ terms of the second one (the function $S$ is 
defined in (\ref{E10})). 
From (\ref{ET2}) it follows that
$
R=e^{S(\eta)}$, 
$C/R=\frac{2S'(\eta)}{\pi \theta_2(0)\theta_3(0)}.
$
We can use (\ref{diffS}) with
the substitution $u\to \eta$ to get
\beq\label{ET5}
R^2+R^{-2}\Bigl (1-\frac{C^2}{4}\Bigr )=
2\cosh \left (2F_{0\bar 0}\right )-
e^{-F_{00}-F_{\bar 0\bar 0}}F_{\bar 0 1}F_{0\bar 1}=
\frac{\theta_2^2(0|\tau)}{\theta_3^2(0|\tau)}+
\frac{\theta_3^2(0|\tau)}{\theta_2^2(0|\tau)}.
\eeq
Similarly to (\ref{E7}), this equation means that
the modular parameter $\tau$ is expressed neatly through
second order partial derivatives of the function $F$.
The same is true for $c_1$ and $\eta$.

Equations (\ref{ET3})
imply the following relations:
\beq\label{}
\begin{array}{l}
\nabla (z_1)S\Bigl (u(z_2)\Bigr )=
\p_{t_0}S\Bigl ( u(z_1)\! -\! u(z_2) \Bigr ),
\;\;
\nabla (z_1)S\Bigl (u(z_2)+\eta \Bigr )=
\p_{\bar t_0}S\Bigl ( u(z_1)\! -\! u(z_2) \Bigr ),
\\ \\
\bar \nabla (\bar z_1)S\Bigl (u(z_2)\Bigr )=
\p_{t_0}S\Bigl ( \bar u(\bar z_1)\! +\! u(z_2)\! +\! \eta \Bigr ),
\;\;
\bar \nabla (\bar z_1)S\Bigl (u(z_2)\! +\! \eta\Bigr )=
\p_{\bar t_0}S\Bigl ( \bar u(\bar z_1)\! +\! u(z_2)\! +\! \eta \Bigr ).
\end{array}
\eeq
In particular, we have $\nabla (z)\log R = \p_{\bar t_0}S(u(z))=
\p_{t_0}S(u(z)+\eta)$.

\section{Comparison with other hierarchies}

It is instructive to compare
the dispersionless 
Pfaff-type hierarchies 
with the more familiar 
dispersionless KP (dKP), mKP (dmKP) and 2DTL (d2DTL) ones:
\beq\label{dkp}
\mbox{dKP:} \qquad e^{D(z)D(\zeta )F}=1-\frac{\p_{t_1}\bigl (D(z)\! -\! 
D(\zeta )\bigr )F}{z-\zeta}\,,
\eeq
\beq\label{dmkp}
\mbox{dmKP:} \qquad e^{D(z)D(\zeta )F}=\frac{ze^{-\p_{t_0}D(z)F}-
\zeta e^{-\p_{t_0}D(\zeta )F}}{z-\zeta}
\,,
\eeq
\beq\label{dtoda}
\mbox{d2DTL:}\quad
\left \{ \begin{array}{l}\displaystyle{
e^{D(z)D(\zeta )F}=\frac{ze^{-\p_{t_0}D(z)F}-
\zeta e^{-\p_{t_0}D(\zeta )F}}{z-\zeta}}
\\ \\
\displaystyle{
e^{-D(z)\bar D(\bar \zeta)F}=1-(z\bar \zeta )^{-1}e^{\p_{t_0}
(\p_{t_0}+D(z)+
\bar D(\bar \zeta))F}}.
\end{array}\right.
\eeq
In the dKP case the (real) times are $\{t_1, t_2, \ldots \}$.
In the dmKP case this set of times is supplemented 
by (real) $t_0$.
In the d2DTL case the time $t_0$ is real
while the other ones are complex, i.e. we have 
two sets of times $\{t_1, t_2, \ldots \}$ and
$\{\bar t_1, \bar t_2, \ldots \}$ which are complex conjugate 
to each other. Note that the first equation in (\ref{dtoda})
and equation (\ref{dmkp}) are identical.
For more details see \cite{TakTak,Tak2014}.

\paragraph{dPfaff-KP versus dKP and dmKP.} First of all, let us
note that the dmKP equation (\ref{dmkp}) implies (\ref{dkp}).
Indeed, writing (\ref{dmkp}) in the form
$z_{12}e^{D_1D_2F}=z_1e^{-D_1F_0}\! -\! z_2e^{-D_2F_0}$
and summing such equations 
for the pairs $12$, $23$, $31$, we get 
$z_{12}e^{D_1D_2F}+z_{23}e^{D_2D_3F}+z_{31}e^{D_1D_3F}=0$,
and tending $z_3\to \infty$ here, we arrive at (\ref{dkp}).
In a similar way, one can show that equation
(\ref{dmkp}), written through the operator $\nabla (z)$ 
in the form $(z_1^{-1}-z_2^{-1})e^{\nabla_1 \nabla_2F}=
z_1^{-1}e^{-\nabla_1F_0}-z_2^{-1}e^{-\nabla_2F_0}$,
implies the equation
\beq\label{dmkp1}
\left |\begin{array}{ccc}
1 & z_1^{-1}& e^{\nabla_2\nabla_3F}
\\ && \\
1 & z_2^{-1}& e^{\nabla_1\nabla_3F}
\\ && \\
1 & z_3^{-1}& e^{\nabla_1\nabla_2F}
\end{array}
\right |=0.
\eeq
In its turn, this equation implies similar antisymmetric determinant 
relations containing more points. In particular, it is an easy
algebraic exercise to show that it follows from (\ref{dmkp1}) that
\beq\label{E99}
\left |
\begin{array}{llll}
1 & z_1^{-1} &z_1^{-2} & e^{(\nabla _2\nabla _3+
\nabla _3\nabla _4+\nabla _4\nabla _2)F}
\\ &&& \\
1 & z_2^{-1} &z_2^{-2} & e^{(\nabla _1\nabla _3+
\nabla _3\nabla _4+\nabla _4\nabla _1)F}
\\ &&& \\
1 & z_3^{-1} &z_3^{-2} & e^{(\nabla _1\nabla _2+
\nabla _2\nabla _4+\nabla _4\nabla _1)F}
\\ &&& \\
1 & z_4^{-1} &z_4^{-2} & e^{(\nabla _1\nabla _2+
\nabla _2\nabla _3+\nabla _3\nabla _1)F}
\end{array}
\right |=0.
\eeq
In fact this is the dispersionless limit of one of the higher 
equations of the difference Hirota hierarchy \cite{Zabrodin97}.

Now let us turn to the dPfaff-KP hierarchy in the elliptic form.
Plugging the left hand side of equation (\ref{E4}) 
(for different pairs of variables)
into the identity
\beq\label{E8}
\begin{array}{ll}
&\displaystyle{\frac{\theta_1(u_{12})\theta_1(u_{23})
\theta_1(u_{31})}{\theta_4(u_{12})\theta_4(u_{23})
\theta_4(u_{31})} \, -\, 
\frac{\theta_1(u_{12})\theta_1(u_{24})
\theta_1(u_{41})}{\theta_4(u_{12})\theta_4(u_{24})
\theta_4(u_{41})}}
\\ &\\
+&\displaystyle{
\frac{\theta_1(u_{13})\theta_1(u_{34})
\theta_1(u_{41})}{\theta_4(u_{13})\theta_4(u_{34})
\theta_4(u_{41})} \, -\, 
\frac{\theta_1(u_{23})\theta_1(u_{34})
\theta_1(u_{42})}{\theta_4(u_{23})\theta_4(u_{34})
\theta_4(u_{42})}\, =\, 0,
}
\end{array}
\eeq
we get the 
determinant relation for the function $F$ of {\it precisely the same 
form (\ref{E99}) as the higher 
equation of the dmKP hierarchy}. At the same time, none of solutions to the 
latter obey equations (\ref{D1}), (\ref{D2}) of the dPfaff-KP 
hierarchy. Indeed, as is shown above, equation (\ref{dkp}) is valid
for the dmKP hierarchy; plugging it into (\ref{D1}),
we get $e^{2\p_{t_0}(2\p_{t_0}+D(z)+D(\zeta))F}=0$ which is impossible
for any $F$. For example, the simplest solution to (\ref{dmkp1})
and (\ref{E99}) 
is $F=0$ which is not a solution to (\ref{D1}).
We see that (\ref{dmkp1}) implies (\ref{E99}) but
not vice versa.

\paragraph{Dispersionless Toda chain.}
Let us also mention the familiar reduction of the d2DTL 
hierarchy obtained by imposing the conditions $\p_{t_k}F=
\p_{\bar t_k}F$ for all $k\geq 1$ which means that 
$D(z)F=\bar D(z)F$. This hierarchy is called the dispersionless
Toda chain (dTC):
\beq\label{dtc}
\mbox{dTC:}\quad
\left \{ \begin{array}{l}\displaystyle{
e^{D(z)D(\zeta )F}=\frac{ze^{-\p_{t_0}D(z)F}-
\zeta e^{-\p_{t_0}D(\zeta )F}}{z-\zeta}}
\\ \\
\displaystyle{
e^{-D(z)D(\zeta)F}=1-(z \zeta )^{-1}e^{\p_{t_0}
(\p_{t_0}+D(z)+ D(\zeta))F}}.
\end{array}\right.
\eeq
In terms of the function $\omega (z)=ze^{-\frac{1}{2}F_{00}
-\p_{t_0}D(z)F}$,
we can rewrite (\ref{dtc}) as
\beq\label{dtc1}
\left \{ \begin{array}{l}\displaystyle{
e^{-\frac{1}{2}F_{00}+D(z)D(\zeta )F}=
\frac{\omega (z)-\omega (\zeta )}{z-\zeta}}
\\ \\
\displaystyle{
e^{-D(z)D(\zeta)F}=1-\frac{1}{\omega (z)\omega (\zeta )}}\,.
\end{array}\right.
\eeq
Note the similarity with (\ref{D1a}), (\ref{D2a}).
Following \cite{Kodama1}, we 
multiply these equations to conclude that the combination
$z-e^{\frac{1}{2}
F_{00}}(\omega (z)+ \omega ^{-1}(z))$ does not depend on $z$.
Tending $z\to \infty$ we find this constant to be equal to $F_{01}$.
Therefore, the variables $z$ and $\omega$ satisfy the algebraic 
equation 
\beq\label{dtc2}
z=e^{\frac{1}{2}F_{00}}\Bigl (
\omega (z)+\frac{1}{\omega (z)}\Bigr )+F_{01},
\eeq
which defines a rational (genus $0$) curve.

Finally, we shall show that the dPfaff-KP hierarchy (\ref{D1}), (\ref{D2})
contains the dispersionless Toda chain as a reduction
(see \cite[Proposition 4.1]{Kodama1}). Consider solutions 
to the dPfaff-KP hierarchy such that $\p_{t_{2k+1}}F=0$ for all
$k\geq 0$. Redefine the times as follows: 
$\tilde t_n=2t_{2n}$, $n\geq 1$, $\tilde t_0 =\frac{1}{2}\, t_0$.
Then it is easy to check that equations 
(\ref{D1}), (\ref{D2}) convert into the system
\beq\label{dtc3}
\left \{ \begin{array}{l}\displaystyle{
e^{\tilde D(z^2)\tilde D(\zeta ^2 )\tilde F}=
\frac{ze^{-\p_{\tilde t_0}\tilde D(z^2 )\tilde F}-
\zeta e^{-\p_{\tilde t_0}\tilde D(\zeta ^2)\tilde F}}{z^2-\zeta ^2}}
\\ \\
\displaystyle{
e^{-\tilde D(z^2 )\tilde D(\zeta ^2)\tilde F}=
1-(z^2 \zeta ^2)^{-1}e^{\p_{\tilde t_0}
(\p_{\tilde t_0}+\tilde D(z^2)+ \tilde D(\zeta ^2))\tilde F}}
\end{array}\right.
\eeq
for the function $\tilde F(\tilde t_0, \tilde t_1, \tilde t_2, 
\ldots )=F(t_0, 0, t_2, 0, t_4, 0, \ldots )$, where 
$\displaystyle{\tilde D(z)=\sum_{k\geq 1}\frac{z^{-k}}{k}
\frac{\p}{\p \tilde t_k}}$, which is equivalent to 
(\ref{dtc}).

\section*{Appendix}

The Jacobi's theta functions $\theta_a (u)=
\theta_a (u|\tau )$, $a=1,2,3,4$, are defined by the formulas
\beq\label{Bp1}
\begin{array}{l}
\theta _1(u)=-\displaystyle{\sum _{k\in \z}}
\exp \left (
\pi i \tau (k+\frac{1}{2})^2 +2\pi i
(u+\frac{1}{2})(k+\frac{1}{2})\right ),
\\
\theta _2(u)=\displaystyle{\sum _{k\in \z}}
\exp \left (
\pi i \tau (k+\frac{1}{2})^2 +2\pi i
u(k+\frac{1}{2})\right ),
\\
\theta _3(u)=\displaystyle{\sum _{k\in \z}}
\exp \left (
\pi i \tau k^2 +2\pi i u k \right ),
\\
\theta _4(u)=\displaystyle{\sum _{k\in \z}}
\exp \left (
\pi i \tau k^2 +2\pi i
(u+\frac{1}{2})k\right ),
\end{array}
\eeq 
where the modular parameter $\tau$ is 
such that ${\rm Im}\, \tau >0$. The function 
$\theta_1(u)$ is odd, the other three functions are even.
It is convenient to 
understand the index $a$ modulo $4$, 
i.e., to identify $\theta_{a} (z) \equiv \theta_{a+4} (z)$. 
Set $ \omega _0 =0$, $\omega_1 =\frac{1}{2}$, 
$\omega _2=\frac{1+\tau}{2}$, $ \omega _3
=\frac{\tau}{2}$ then the function $\theta _a(u)$ 
has simple zeros at the points of the lattice $\omega _{a-1}+\ZZ +\ZZ
\tau $. The theta functions have the following 
quasi-periodic properties under shifts
by $1$ and $\tau$: 
\beq\label{Bp1a}
\begin{array}{l}
\theta _a (u+1)=e^{\pi i(1+2\p_{\tau}\omega_{a\! -\! 1})}
\theta _a(u),
\\
\theta _a (u+\tau )=e^{\pi i(a+2\p_{\tau}\omega_{a\! -\! 1})}
e^{-\pi i\tau -2\pi iu}\theta _a(u).
\end{array}
\eeq
Shifts by the half-periods relate
the different theta functions to each other.
We also mention the identity
\beq\label{theta1prime}
\theta_1'(0)=\pi \theta_2(0) \theta_3(0) \theta_4(0).
\eeq
Many useful formulas with the theta functions can be found 
in \cite{KZtheta}.

\section*{Acknowledgements}


We thank S.Kharchev, I.Krichever, S.Natanzon, 
A.Orlov and T.Takebe for discussions.
The authors were supported in part by RFBR grant
14-02-00627. The work of A.Z. was also partially supported 
by joint RFBR grant 14-01-90405-Ukr and by grant NSh-1500.2014.2 
for support of leading scientific schools.
The financial support from the Government of the Russian Federation 
within the framework of the implementation of the 5-100 Programme Roadmap 
of the National Research University  Higher School of Economics is acknowledged.
Some results of the paper were reported at the
4th Workshop on combinatorics of moduli spaces,
cluster algebras, and topological recursion (Moscow, May 26-31, 2014).

\end{document}